\documentclass[aps,prd,onecolumn,groupedaddress,preprintnumbers,superscriptaddress,
showpacs,nofootinbib]{revtex4}
\usepackage{amsmath,amssymb,latexsym,mathrsfs}
\usepackage{color}

\begin{document}

\title{
Late-time symmetry near black hole horizons
}

\author{Kentaro Tanabe}
\affiliation{Yukawa Institute for Theoretical Physics, Kyoto University, Kyoto 606-8502, Japan}

\author{Tetsuya Shiromizu}
\author{Shunichiro Kinoshita}
\affiliation{Department of Physics, Kyoto University, Kyoto 606-8502, Japan}

\preprint{YITP-11-99} 

\begin{abstract}
It is expected that black holes are formed dynamically under gravitational collapses 
and approach stationary states. In this paper, we show that the asymptotic Killing 
vector at late time should exist on the horizon and then that it can be extended outside 
black holes under the assumption of the analyticity of spacetimes. This fact implies that 
if there is another asymptotic 
Killing vector which becomes a stationary Killing at a far region and spacelike in the ``ergoregion,'' 
the rotating black holes may have the asymptotically axisymmetric Killing vector at late time. Thus, we may expect 
that the asymptotic rigidity of the black holes holds.
\end{abstract}


\maketitle

\section{introduction}

Black holes in our Universe are expected to be formed under gravitational collapses, 
and to finally approach stationary
 and vacuum states by radiating and absorbing energy, 
momentum, and angular momentum. Then the uniqueness theorem~\cite{Israel} guarantees that the 
black hole candidates in our Universe are the Kerr black holes. The key ingredient for the proof of the uniqueness theorem is the 
rigidity theorem~\cite{Hawking:1973uf, Hawking:1971vc, Friedrich:1998wq, Racz:1999ne}. 
The rigidity theorem shows that the stationary rotating black holes have axisymmetric Killing vectors. 
The outline of the proof is as follows: The stationarity of spacetime implies that there 
are no gravitational waves around a black hole. Then the expansion and shear of the event horizon 
must vanish by virtue of the Raychaudhuri equation and stationarity. Using the Einstein equations, then, 
we can find that the null geodesic generator of the event horizon is a Killing vector. 
If the black hole is rotating, this new Killing vector may deviate from the stationary 
Killing vector which becomes spacelike in the ergoregion. 
This means that the stationary black holes might have two Killing vectors, 
that is, the stationary and axisymmetric Killing vectors. 
Hence the stationary black holes should rotate rigidly.    
 
However, the late-time phase of black holes produced by gravitational collapses would not be 
exactly stationary but nearly stationary. ``Nearly stationary'' means that the black holes 
are surrounded by the gravitational waves at late time. Then we cannot apply the rigidity 
theorem to such black holes because the expansion and shear of the null geodesic generator 
on the event horizon do not vanish due to the presence of the gravitational waves 
on the event horizon. Note that the late-time behaviors of the perturbations around the 
Schwarzschild and Kerr black holes were examined and it was shown that the perturbations 
decay at late time both on the horizon and null 
infinity at the same rate (for examples, see Refs.~\cite{Gundlach:1993tp, Barack:1999ma, Barack:1999ya}). 
In the dynamical processes in gravitational collapses, however, 
it is quite nontrivial whether the formed black hole approaches the Kerr black hole. 
In this paper, we show that the asymptotic Killing vector, which will asymptotically approach a Killing vector at late time, 
should exist on the horizon without assuming any symmetries 
and then that it can be extended outside of the event horizon using the Einstein equations. 
This indicates that if there is 
another asymptotic Killing vector which will be an asymptotically stationary Killing vector 
at a far region, the rotating black hole may have the asymptotically axisymmetric Killing 
vector at late time.

This paper is organized as follows. In the next section we provide the Einstein equations 
near the future event horizons. In Sec. III, we investigate the late-time behaviors of the metric 
on the event horizon and find the late-time symmetry. Then, under the assumption of the analyticity, 
we show that this late-time symmetry can be extended outside of the event horizon using the Einstein 
equation. In Sec. IV, we provide a summary and discussion. In Appendix A we perform the $(n+1)$ decompositions 
of the Einstein equation. Using them, we provide the explicit form of the 
Einstein equation in the current form of the metric in Appendix B. Almost a similar formulation 
is developed in the study on null infinity~\cite{Tanabe2011}. 
In Appendix C, we present the details of the discussion associated with 
gauge freedom.

\section{Bondi-like coordinate and Einstein equations}

In this section, introducing the Bondi-like coordinate near the event horizon, 
we investigate the initial value problem. For the details of these derivations, 
see Appendixes A and B.

\subsection{Bondi-like coordinate near event horizon}

We consider a dynamical black hole and investigate the late-time behavior of the near horizon 
geometry at late time. The black hole is defined as the region which is not contained 
in $J^{-}(\mathscr{I}^{+})$ where $\mathscr{I}^{+}$ is the future 
null infinity. $J^-(S)$ is the chronological past connecting a set $S$ by causal curves from $S$. 
The event horizon is the boundary of the black hole and it is a null hypersurface. See Ref.~\cite{Hawking:1973uf} for the precise definitions. 
Then, we introduce the Bondi-like (Gaussian null) coordinates $x^{A}=(u,r,x^{a})$ 
near the event horizon as
%============<Equation>=============%
%
\begin{equation}
ds^{2}=g_{AB}dx^Adx^B=-Adu^{2}+2dudr+h_{ab}(dx^{a}+U^{a}du)(dx^{b}+U^{b}du),
\end{equation} 
%
%===================================%
where $u$ is a time coordinate. In this coordinate, the horizon position is taken to be $r=0$. 
We assume that a cross section of the event horizon is compact in the $u=\text{constant}$ hypersurface and its topology is $S^{2}$. 
$x^{a}$ are coordinates 
on $S^{2}$. Since the event horizon is a null hypersurface, $g_{uu}$
must vanish on the event horizon and $l=\partial/\partial u$ is the null
geodesic generator on the event horizon.
Furthermore we can choose the coordinate $u$ so that $l_{a}=0$ on the
event horizon.
Then, we have 
%============<Equation>=============%
%
\begin{equation}
A\,\hat{=}\,0, \label{A-cond}
\end{equation} 
%
%===================================%
and 
%============<Equation>=============%
%
\begin{equation}
U^{a}\,\hat{=}\,0, \label{Ua-cond}
\end{equation} 
%
%===================================%
where $\hat{=}$ means the evaluation on the event horizon $r=0$. 

To solve the vacuum Einstein equations, we formulate the initial value problem in the Bondi coordinates. 
For the convenience of our study in the following sections, we solve the Einstein equations 
as the evolution equations in the direction of $r$. Thus, the initial value of the metric should be set 
on the $r=\text{constant}$ surface. In this paper we take the event horizon $r=0$ as the initial surface. 

The evolution equations are given by     
%============<Equation>=============%
%
\begin{eqnarray}
R_{rr}&=&
-\frac{1}{2}(\log h)'' -\frac{1}{4} h^{ac}h^{bd}(h_{ab})^{'}(h_{cd})^{'} \,=\,0,  \label{rr-eq} 
\end{eqnarray}
\begin{eqnarray}
R_{rB}h^{aB}&=&
\frac{1}{2}U^{a''}
+\frac{1}{2}h^{ac}h_{bc}^{'}U^{b'}+\frac{1}{4}(\log h)'U^{a'}
   +\frac{1}{2}h^{ab}\bar{D}^{c}[h_{bc}^{'} - h_{bc}(\log h)']\,=\,0 \label{ra-eq}
\end{eqnarray}
and
\begin{eqnarray}
R_{AB}h^{A}_{~a}h^{B}_{~b}&=&-\frac{1}{2}Ah^{''}_{ab}-\frac{1}{2}A^{'}h^{'}_{ab}
- (\dot h_{ab})' +
{}^{(h)}R_{ab}+\frac{A}{2}h^{cd}h^{'}_{ac}h^{'}_{bd}+\frac{1}{2}h^{cd}(h^{'}_{ac}\dot{h}_{bd}+h^{'}_{bd}\dot{h}_{ac}) 
+ \mathcal L_U h'_{ab}\notag \\
&&~~-\frac{1}{2}h^{'}_{ac}(\bar{D}_{b}U^{c}+\bar{D}^{c}U_{b})-\frac{1}{2}h^{'}_{bc}(\bar{D}_{a}U^{c}+\bar{D}^{c}U_{a}) 
-\frac{1}{2}h_{ac}h_{bd}U^{c}{}^{'}U^{d}{}^{'}
 -\frac{1}{4}[\dot{(\log{h})} - 2\bar{D}_{a}U^{a}]h^{'}_{ab} \notag \\
&&~~-\frac{1}{4}(\log{h})^{'}
(A h^{'}_{ab} +\dot{h}_{ab} -\bar{D}_{a}U_{b}-\bar{D}_{b}U_{a}) 
+\frac{1}{2}(h_{bc}\bar{D}_{a}U^{c}{}^{'}+h_{ac}\bar{D}_{b}U^{c}{}^{'})\,=\,0 \label{ab-eq},
\end{eqnarray} 
%
%===================================%
where the prime and dot denote the $r$ and $u$ derivative, respectively, $\bar{D}_{a}$ is a covariant 
derivative with $h_{ab}$ and $h=\det{h_{ab}}$.
Also, ${}^{(h)}R_{ab}$ is the Ricci tensor with respect to $h_{ab}$.
The evolutions of the metric functions $A$, $U^{a}$ and $h_{ab}$ are
determined by Eqs.~(\ref{rr-eq}), (\ref{ra-eq}) and (\ref{ab-eq}) completely. 
Note that the trace part of Eq.~(\ref{ab-eq}) gives us 
%============<Equation>=============%
%
\begin{eqnarray}
A^{'}(\log{h})^{'}&=&2{}^{(h)}R-\frac{A}{2}[(\log{h})^{'}]^{2}-[\dot{(\log{h})}-2\bar{D}_{a}U^{a}](\log{h})^{'}-h_{ab}U^{a'}U^{b'}
\notag \\
&&~~~~~~~~~~~~~~~~~~~~~~~~~~~~~~~~~~~~~~~~~~~~~~~
-A(\log{h})^{''}-2\dot{(\log{h})}{}^{'}+2(\bar{D}_{a}U^{a})^{'} 
+ U^a \bar{D}_a (\log h)'
\label{aa-eq}, 
\end{eqnarray} 
%
%===================================%
which is used as the evolution equation for $A$.

The other components of the Einstein equations, $R_{uu}=0$, $R_{ur}=0$ and
$R_{ua}=0$, are related to the above evolution equations by the
Bianchi identities.
Therefore, once they are satisfied at the initial surface $r=0$, we do
not need to solve them any more.
In fact, if the evolution equations are satisfied, the Bianchi identities
lead to 
\begin{equation}
 \left\{
  \begin{aligned}
   (\sqrt{h}R_{ua})' &=\, \sqrt{h} R_{ur},\\
   (\sqrt{h}R^r{}_u)' &=\, - \sqrt{h} \bar D_a R^a{}_u 
   - \dot{(\sqrt{h})} R_{ur},\\
   (\log h)' R_{ur} &=\, 0,
  \end{aligned}
 \right.
\end{equation}
where $R^r{}_u = R_{uu} + A R_{ur} - U^a R_{ua}$ and 
$R^a{}_u = - U^a R_{ur} + h^{ab}R_{ub}$.
Thus, the evolution equations guarantee that $R_{uu}=0$, $R_{ur}=0$ and
$R_{ua}=0$ will always be satisfied at $r \ne 0$ if 
$R_{uu} \,\hat{=}\, 0$, $R_{ur} \,\hat{=}\, 0$
and $R_{ua} \,\hat{=}\, 0$ at the initial surface $r=0$.
For convenience, at $r=0$ we will use the following constraint equations: 
\begin{equation}
 R_{uu} \hat{=} - \frac{1}{2}\ddot{(\log h)} + \frac{A'}{4}\dot{(\log h)}
  - \frac{1}{4}h^{ac}h^{bd}\dot h_{ab} \dot h_{cd} = 0, \label{uu-const}
\end{equation}
and 
\begin{equation}
 R^{ra} \hat{=} - \frac{1}{2}(\dot U^a)'
  - \frac{1}{4}\dot{(\log h)}{U^a}' - \frac{1}{2}h^{ab}{U^c}' \dot h_{bc}
  - \frac{1}{2} \bar D^a A' + \frac{1}{2}h^{ab}\bar D^c \dot h_{bc}
  - \frac{1}{2}\bar D^a \dot{(\log h)} = 0 \label{ra-const}
\end{equation}
[see Eqs.~(\ref{uu-eq}) and (\ref{ra-eq2})]. 
Moreover, since $A$ should vanish on the initial surface $r=0$ 
[see Eq.~(\ref{A-cond})], the evolution equations Eq.~(\ref{ab-eq}) become
singular on $r=0$.
Analyticity of $h_{ab}$ on $r=0$ gives us the regularity condition 
%============<Equation>=============%
%
\begin{eqnarray}
 -\frac{1}{2}A^{'}h^{'}_{ab}&-&\dot{h}^{'}_{ab}+
  {}^{(h)}R_{ab}
  + \frac{1}{2}h^{cd}(h^{'}_{ac}\dot{h}_{bd}+h^{'}_{bd}\dot{h}_{ac})
  \notag \\
 &-&
  \frac{1}{2}h_{ac}h_{bd}U^{c}{}^{'}U^{d}{}^{'}
  -\frac{1}{4}\dot{(\log{h})}h^{'}_{ab} \notag \\
 &-&\frac{1}{4}(\log{h})^{'}
  \dot{h}_{ab}
  +\frac{1}{2}(h_{bc}\bar{D}_{a}U^{c}{}^{'}+h_{ac}\bar{D}_{b}U^{c}{}^{'})\,\hat{=}\,0. \label{reg-eq}
\end{eqnarray} 
%
%===================================%

If we give $h_{ab}|_{r=0}$ on $r=0$, we can determine
$h'_{ab}|_{r=0}$, ${U^a}'|_{r=0}$, and $A'|_{r=0}$ by solving
Eqs.~(\ref{uu-const}), (\ref{ra-const}) and (\ref{reg-eq}).
As a result, we will solve the evolution equations~(\ref{rr-eq}),
(\ref{ra-eq}) and (\ref{ab-eq}) using the above initial values on $r=0$.

\subsection{Some explicit solutions near event horizon}

In this subsection, it is shown that we explicitly solve the constraint
equations and evolution equations near the event horizon using power
series expansion around $r=0$.
To do this, 
we expand the metric functions with $r$ near the event horizon as 
%============<Equation>=============%
%
\begin{equation}
A\,=\, rA^{(1)}+\sum_{i\geq 2}r^{i}A^{(i)}, \label{a-exp}
\end{equation}
\begin{equation}
U^{a}\,=\, rU^{(1)a}+ \sum_{i\geq 2}r^{i}U^{(i)a}, \label{u-exp} 
\end{equation}
and
\begin{equation}
h_{ab}\,=\,h^{(0)}_{ab}+rh^{(1)}_{ab}+\sum_{i\geq 2}r^{i}h^{(i)}_{ab}, \label{h-exp}
\end{equation} 
%
%===================================% 
where from the gauge conditions Eqs. (\ref{A-cond}) and (\ref{Ua-cond}), the expansions of $A$ and $U^{a}$ 
start from the first order of $r$. In the following, the tensor index of $h^{(i)}_{ab}$ and $U^{(i)a}$ is raised and lowered 
by $h^{(0)}_{ab}$. The trace and traceless parts are also taken by $h^{(0)}_{ab}$. 

First let us solve the constraint equations and regularity conditions
in order to determine initial values. 
On the initial surface $r=0$, 
$h^{(0)}_{ab}$, $h^{(1)}_{ab}$, $A^{(1)}$ and $U^{(1)a}$ should be set on $r=0$ as initial values. The constraint equations for 
these initial values are $R_{uu}\,\hat{=}\,0$ and $R^{ra}\,\hat{=}\,0$. 
Now $R_{uu}\,\hat{=}\,0$ [Eq.~(\ref{uu-const})] is rewritten as 
%============<Equation>=============%
%
\begin{equation}
\ddot{h}^{(0)}-\frac{1}{2}A^{(1)}\dot{h}^{(0)}
+ \frac{1}{2} h^{(0)ac}h^{(0)bd}\dot{h}_{ab}^{(0)}\dot{h}_{cd}^{(0)}\,=\,0, 
\end{equation} 
%
%===================================%
where $h^{(0)}=\det{h^{(0)}_{ab}}$.
Thus, $A^{(1)}$ should be given to satisfy
this equation for given $h^{(0)}_{ab}$. 
Also, $R^{ra}\,\hat{=}\,0$ [Eq.~(\ref{ra-const})] becomes 
\begin{equation}
\dot{U}^{(1)a}\,=\,-D^{a}A^{(1)}+h^{(0)ac}D^{b}\dot{h}_{bc}^{(0)}
-D^{a}\dot{(\log h^{(0)})}
-h^{(0)ac}U^{(1)b}\dot{h}^{(0)}_{bc}
-\frac{1}{2}U^{(1)a}\dot{(\log h^{(0)})},
\label{u1-ev}
\end{equation} 
%
%===================================%
where $D_{a}$ is the covariant derivative with respect to
$h^{(0)}_{ab}$.
Then the initial value $U^{(1)a}$ is given to satisfy the above.
The regularity condition Eq.~(\ref{reg-eq}) becomes
%============<Equation>=============%
%
\begin{eqnarray}
-\dot{h}^{(1)}_{ab}&-&\frac{1}{2}A^{(1)}h^{(1)}_{ab}+{}^{(h^{(0)})}R_{ab}+\frac{1}{2}h^{(0)cd}(h^{(1)}_{ac}\dot{h}_{bd}^{(0)}
+h^{(1)}_{bd}\dot{h}_{ac}^{(0)})-\frac{1}{2}h^{(0)}_{ac}h^{(0)}_{bd}U^{(1)c}U^{(1)d}\notag \\
&&~~~~~~~~~~~~~~~~~~~~~~~~~~~~~~~~~~~~~~~
-\frac{1}{4}\dot{(\log{h^{(0)}})}h^{(1)}_{ab}
-\frac{1}{4}h^{(0)cd}h^{(1)}_{cd}\dot{h}^{(0)}_{ab}
+\frac{1}{2}(D_{a}U^{(1)}_{b}+D_{b}U^{(1)}_{a})=0, \label{ab-const}
\end{eqnarray} 
%
%===================================%
where ${}^{(h^{(0)})}R_{ab}$ is the Ricci tensor of $h^{(0)}_{ab}$.
We obtain $h^{(1)}_{ab}$ satisfying this equation.
Hence, if we give $h^{(0)}_{ab}$ on the initial surface, the constraint
equations and the regularity condition yield $h^{(1)}_{ab}$, $A^{(1)}$
and $U^{(1)a}$.

Next we solve the evolution equations. Equation (\ref{rr-eq}) becomes near the event horizon as
%============<Equation>=============%
%
\begin{equation}
R_{rr}\,=\,- h^{(0)ab}h^{(2)}_{ab} + \frac{1}{4}(h^{(1)}_{ab})^{2} +O(r)\,=\,0.
\end{equation}%
%===================================%
This equation gives us the trace part of $h^{(2)}_{ab}$ as 
%============<Equation>=============%
%
\begin{equation}
h^{(0)ab}h^{(2)}_{ab}\,=\,\frac{1}{4}(h^{(1)}_{ab})^{2}. \label{h0h2h1}
\end{equation}
%
%===================================% 
The evolution equation of $U^{a}$ as $R_{rB}h^{aB}=0$ [Eq.~(\ref{ra-eq})] can be expanded as
%============<Equation>=============%
%
\begin{equation}
R_{rB}h^{aB}\,=\, U^{(2)a} +
 \frac{1}{2}h^{(0)ac}h^{(1)}_{bc}U^{(1)b} + 
 \frac{1}{4}U^{(1)a}h^{(0)bc}h^{(1)}_{bc}
 + \frac{1}{2}h^{(1)ac}D^{b}\left( h^{(1)}_{bc}-h_{bc}^{(0)}h^{(0)de}h^{(1)}_{de}\right) +O(r). 
\end{equation}
%
%===================================%
Then $U^{(2)a}$ is given by 
%============<Equation>=============%
%
\begin{equation}
U^{(2)a}\,=\, - \frac{1}{2}h^{(0)ac}h^{(1)}_{bc}U^{(1)b}
 - \frac{1}{4}U^{(1)a}h^{(0)bc}h^{(1)}_{bc}
-\frac{1}{2}h^{(1)ac}D^{b}\left( h^{(1)}_{bc}-h_{bc}^{(0)}h^{(0)de}h^{(1)}_{de}\right).\label{u2-f}
\end{equation}
%
%===================================%
In the same way, expanding the evolution equations Eq.~(\ref{ab-eq})
near the event horizon, we can obtain the traceless part
$h^{(2)}_{\langle ab\rangle}$ and $A^{(2)}$ in terms of
$h^{(0)}_{ab}, h^{(1)}_{ab}, U^{(1)a}$ and $A^{(1)}$.
Note that $A^{(2)}$ is given by the trace part of Eq. (\ref{ab-eq}), namely Eq.~(\ref{aa-eq}).
However, we do not provide its
explicit form because its form is very cumbersome. 

Hence, we can determine $h^{(2)}_{ab}$, $U^{(2)a}$, and $A^{(2)}$ using
the evolution equations.
To determine the higher order quantities, $h^{(i)}_{ab}$, $U^{(i)a}$, and $A^{(i)}$
($i>2$), we have to repeat the same procedure.

\section{Late-time symmetry on and near event horizon}

In this section, we show that there is late-time symmetry on the event horizon. Then we 
will extend it outside of black hole regions. 

\subsection{Late-time behaviors on event horizon}

To investigate late-time behaviors of the event horizon, we introduce the expansion and shear of the null geodesic generator 
$l=\partial/\partial u$ of the event horizon. The expansion $\theta$ and shear $\sigma_{ab}$ are defined as
%============<Equation>=============%
%
\begin{eqnarray}
\sigma_{ab} +\frac{1}{2}\theta h^{(0)}_{ab}&\hat{=}&h_{a}^{~A}h_{b}^{~B}\nabla_{A}l_{B} \notag \\
&\hat{=}&\frac{1}{2}\dot{h}^{(0)}_{ab}, 
\end{eqnarray} 
%
%===================================% 
where $\sigma_{ab}$ is  the traceless part of $\dot{h}_{ab}^{(0)}$ with respect to $h^{(0)}_{ab}$.
Then we can rewrite Eq.~(\ref{uu-const}), one of the constraint
equations, 
using $\theta$ and $\sigma_{ab}$ as 
%============<Equation>=============%
%
\begin{equation}
\dot{\theta}-\frac{1}{2}A^{(1)}\theta\,=\,-\frac{1}{2}\theta^{2} -\sigma_{ab}\sigma^{ab}. \label{ray-eq1}
\end{equation} 
%
%===================================% 

We can regard $A^{(1)}/2$ as a surface gravity of the black hole with respect to the time coordinate $u$ because
%============<Equation>=============%
%
\begin{equation}
l^{A}\nabla_{A}l^{B}\,\hat{=}\,\frac{A^{(1)}}{2}l^{B}
\end{equation} 
%
%===================================% 
holds. Using the affine parameter $w$ defined by
%============<Equation>=============%
%
\begin{equation}
\frac{dw}{du}\,=\,\exp \Bigl( \int^{u}\!\frac{A^{(1)}}{2}du' \Bigr),
\end{equation} 
%
%===================================%
we can obtain the Raychaudhuri equation
%============<Equation>=============%
%
\begin{equation}
\partial_{w}\theta_{(w)}\,=\,-\frac{1}{2}\theta_{(w)}^{2}-\sigma_{(w)ab}\sigma_{(w)}^{ab}, \label{ray-eq2} 
\end{equation} 
%
%===================================%
where $\theta_{(w)}$ and $\sigma_{(w)ab}$ are expansion and shear with respect to $w$. We can see 
the relation between $\theta$, $\sigma_{ab}$ and $\theta_{(w)}$, $\sigma_{(w)ab}$ as 
%============<Equation>=============%
%
\begin{equation}
\theta\,=\,\theta_{(w)} \exp{\Bigl( \int^{u}\!\frac{A^{(1)}}{2}du' \Bigr)}\,,\quad 
\sigma_{ab}\,=\,\sigma_{(w)ab} \exp{\Bigl( \int^{u}\!\frac{A^{(1)}}{2}du' \Bigr)}.
\end{equation} 
%
%===================================% 
Here we remember that the area law of the event horizon holds for spacetimes satisfying the 
null energy condition, that is, $\theta\geq 0$ and $\theta_{(w)}\geq 0$. 
Since $\sigma_{(w)ab}\sigma_{(w)}^{ab} \geq 0$, Eq.~(\ref{ray-eq2}) implies the 
inequality 
%============<Equation>=============%
%
\begin{equation}
\partial_w \theta_{(w)}+\frac{1}{2}\theta_{(w)}^2 \leq 0.
\end{equation} 
%
%===================================% 
Then the integration over $w$ gives us 
%============<Equation>=============%
%
\begin{equation}
\theta_{(w)} \leq \frac{1}{1/\theta_{(0)}+(w-w_0)/2} \to 0 ~~({\text {as}}~~w \to \infty),
\end{equation} 
%
%===================================% 
where we used the fact of $\theta_{(0)}=\theta_{(w)}(w=w_0) \geq 0$. 
In addition, 
Eq.~(\ref{ray-eq2}) shows that 
the shear $\sigma_{(w)ab}$ should also vanish as $w\to\infty$. 
This is shown as a part of the proof of another theorem \cite{HSN}. 
From now on, we assume that $w\to\infty$ corresponds to $u\to\infty$. Then, we see that 
$\theta$ and $\sigma_{ab}$ should also vanish as $u\to\infty$. 
It is natural to assume that the cross section of the event horizon is compact. 
Then the vanishing of the expansion implies that the horizon area approaches  
a constant and finite value.

Altogether we see the behavior of the metric at late time as
%============<Equation>=============%
%
\begin{equation}
\mathcal{L}_{l}g_{AB}|_{\rm horizon} \to 0~~(u\to\infty).
\end{equation} 
%
%===================================%   

Here we impose the following decaying condition on the event horizon for the metric:
%============<Equation>=============%
%
\begin{equation}
\mathcal{L}_{l}g_{AB}\,\hat{=}\,O\left( \frac{1}{u^{n}}\right), \label{asBH}
\end{equation} 
%
%===================================%   
which explicitly means $\dot{h}^{(0)}_{ab} = O(u^{-n})$.
This equation means that the null geodesic generator of the event horizon $l$ should be an asymptotic 
Killing vector at late time ($u\rightarrow\infty$). Thus there is a late-time symmetry on the event horizon.  

Since we consider the dynamics only near the event horizon, we cannot 
determine the decaying rate of the metric. However, the details of the decaying properties 
are not important for our argument, that is, our result does not depend on $n$. 
Note that it will be determined by the boundary conditions at a far region from the black holes as 
in Refs. \cite{Gundlach:1993tp, Barack:1999ma, Barack:1999ya}. 

It should be remembered that the horizon which satisfies this condition is called 
a slowly evolving horizon in Refs. \cite{Booth:2003ji, Booth:2006bn}. 
If the right-hand side of Eq.~(\ref{asBH}) vanishes, the event horizon  will 
be identical to the isolated horizon \cite{Ashtekar:2004cn}. 

\subsection{Extension of late-time symmetry}

The purpose of this subsection is to show that the decaying condition of Eq.~(\ref{asBH}) 
can be extended outside of the event horizon as 
%============<Equation>=============%
%
\begin{equation}
\mathcal{L}_{l}g_{AB}\,=\,O\left( \frac{1}{u^{n}}\right). \label{rigid}
\end{equation} 
%
%===================================%
In the following we assume the analyticity of $g_{AB}$. 

Under the presence of the analyticity of spacetimes, the above is equivalent with 
%============<Equation>=============%
%
\begin{equation}
(\mathcal{L}_{n})^{m}\mathcal{L}_{l}g_{AB}\,\hat{=}\,O\left( \frac{1}{u^{n}}\right), \label{rigid2}
\end{equation} 
%
%===================================%
where $n=\partial/\partial r$ and $m=0,1,2,\cdots$. Note that ``$\hat{=}$'' means the evaluation 
on the event horizon $(r=0)$ again. 

First we will show the $m=1$ case: 
%============<Equation>=============%
%
\begin{equation}
\mathcal{L}_{n}\mathcal{L}_{l}g_{AB}\,\hat{=}\,O\left( \frac{1}{u^{n}}\right). \label{nl}
\end{equation} 
%
%===================================%
Substituting the explicit form of $g_{AB}$ to the above, we rewrite Eq.~(\ref{nl}) as 
%============<Equation>=============%
%
\begin{equation}
-\mathcal{L}_{n}\mathcal{L}_{l}(A-U_{a}U^{a}) (du)_{A}(du)_{B}
 + 2\mathcal{L}_{n}\mathcal{L}_{l}U_{a}(du)_{(A}(dx^{a})_{B)}
+\mathcal{L}_{n}\mathcal{L}_{l}h_{ab}(dx^{a})_{A}(dx^{b})_{B}\hat{=}\,O\left( \frac{1}{u^{n}}\right). \label{ar1}
\end{equation} 
%
%===================================%
Using Eqs.~(\ref{a-exp}), (\ref{u-exp}) and (\ref{h-exp}), the above will be equivalent with  
%============<Equation>=============%
%
\begin{equation}
\mathcal{L}_{l}A^{(1)}\,=\,O\left( \frac{1}{u^{n}}\right), \label{a1-eq}
\end{equation} 
\begin{equation}
\mathcal{L}_{l}U^{(1)a}\,=\,O\left( \frac{1}{u^{n}}\right), \label{u1-eq}
\end{equation} 
and
\begin{equation}
\mathcal{L}_{l}h^{(1)}_{ab}\,=\,O\left( \frac{1}{u^{n}}\right). \label{h1-eq}
\end{equation} 
%
%===================================%
Let us examine these conditions. First, we focus on Eq.~(\ref{a1-eq}). As a result, 
using the gauge freedom, we can show that the slightly strong condition, 
$\mathcal{L}_{l}A^{(1)}\,=\,O\ (1/u^{n+1})$, holds. 
To see this, we decompose $A^{(1)}$ into the $u$-independent term and others as
%============<Equation>=============%
%
\begin{equation}
A^{(1)}\,=\,A^{(1)}_{0}(x^{a})+\tilde{A}^{(1)}(u,x^{a}) \label{a1-eq2}.
\end{equation} 
%
%===================================%
As shown in Ref. \cite{Hollands:2006rj}, we can choose the gauge so that $A^{(1)}_{0}$ is a constant. 
Furthermore, using the residual gauge, we can take $\tilde{A}^{(1)}=O(1/u^{n})$. Then, using the gauge 
freedom in our coordinate, we can make $A^{(1)}$ satisfy stronger decaying property as 
%============<Equation>=============%
%
\begin{equation}
\mathcal{L}_{l}A^{(1)}\,=\,O\left( \frac{1}{u^{n+1}}\right). \label{a1-eq3}
\end{equation} 
%
%===================================%
For the details, see Appendix C. Of course, Eq.~(\ref{a1-eq}) is satisfied. 

Now $U^{(1)a}$ should satisfy the constraint equation of Eq.~(\ref{u1-ev}) as
%============<Equation>=============%
%
\begin{equation}
\mathcal L_l U^{(1)a}\,=\,-D^{a}\tilde{A}^{(1)}+h^{(0)ac}D^{b}\dot{h}_{bc}^{(0)}
-D^{a}\dot{(\log h^{(0)})}
-h^{(0)ac}U^{(1)b}\dot{h}^{(0)}_{bc}
-\frac{1}{2}U^{(1)a}\dot{(\log h^{(0)})}
 .
\label{u1-ev2}
\end{equation}
%
%===================================%
Together with Eqs.~(\ref{asBH}) and (\ref{a1-eq3}), we can see that Eq.~(\ref{u1-eq}) holds from 
the above. 

Furthermore, $h^{(1)}_{ab}$ satisfy the following equation [see Eq.~(\ref{ab-const})] as a regularity condition
%============<Equation>=============%
%
\begin{eqnarray}
\dot{h}^{(1)}_{ab}+\frac{1}{2}A^{(1)}h^{(1)}_{ab}&=&{}^{(h^{(0)})}R_{ab}+\frac{1}{2}h^{(0)cd}(h^{(1)}_{ac}\dot{h}_{bd}^{(0)}
+h^{(1)}_{bd}\dot{h}_{ac}^{(0)})-\frac{1}{2}h^{(0)}_{ac}h^{(0)}_{bd}U^{(1)c}U^{(1)d}\notag \\
&&~~-\frac{1}{4}\dot{(\log{h^{(0)}})}
h^{(1)}_{ab}-\frac{1}{4}h^{(0)cd}h^{(1)}_{cd}\dot{h}^{(0)}_{ab}
+ \frac{1}{2}(D_{a}U^{(1)}_{b}+D_{b}U^{(1)}_{a}), \label{h1-ev}
\end{eqnarray} 
%
%===================================%
Then multiplying $\mathcal{L}_{l}$ to the above and using Eq.~(\ref{u1-ev}), we can see 
%============<Equation>=============%
%
\begin{equation}
\frac{1}{2}A^{(1)} \mathcal{L}_{l}h^{(1)}_{ab} = \mathcal{L}_{l}{}^{(h^{(0)})}R_{ab}+O(u^{-n})
\end{equation} 
%
%===================================%
holds. In the above, the higher-order terms like $\mathcal{L}_{l}^2{h}^{(1)}_{ab}, (\mathcal{L}_{l}h^{(0)}_{ab})^{2}, 
(\mathcal{L}_{l}h^{(1)}_{ab})^{2}$, $\mathcal{L}_{l}h^{(0)}_{ab} \mathcal{L}_{l}h^{(1)}_{ab}$ and so on are 
contained in $O(u^{-n})$. 
Thus Eq.~(\ref{asBH}) tells us that $\mathcal{L}_{l}h^{(1)}_{ab} = O(u^{-n})$ holds. 
As a consequence, we can show the $m=1$ case of Eq.~(\ref{rigid2}).

For the $m>1$ cases of Eq.~(\ref{rigid2}), we perform the same procedure inductively. Let $m>1$ 
be an integer and assume that the metric satisfies 
%============<Equation>=============%
%
\begin{equation}
(\mathcal{L}_{n})^{k}\mathcal{L}_{l}g_{AB}\,\hat{=}O\left( \frac{1}{u^{n}}\right)
\end{equation} 
%
%===================================%
for all $k(<m)$. Then $(\mathcal{L}_{n})^{m}\mathcal{L}_{l}g_{AB}\,\hat{=}O(u^{-n})$ 
is equivalent with 
%============<Equation>=============%
%
\begin{gather}
\mathcal{L}_{l}A^{(m)}\,=\,O\left( \frac{1}{u^{n}}\right), \label{am-eq} \\
\mathcal{L}_{l}U^{(m)a}\,=\,O\left( \frac{1}{u^{n}}\right), \label{um-eq}
\end{gather} 
and
\begin{equation}
\mathcal{L}_{l}h^{(m)}_{ab}\,=\,O\left( \frac{1}{u^{n}}\right). \label{hm-eq}
\end{equation} 
%
%===================================%

Since $U^{(m)a}$ is written by $U^{(1)a}$, $A^{(1)}, h^{(0)}_{ab},\cdots, h^{(m-1)}_{ab}$ 
from Eq.~(\ref{ra-eq}) like Eq.~(\ref{u2-f}), the induction assumption
%============<Equation>=============%
%
\begin{equation}
\mathcal{L}_{l}h^{(i)}_{ab}=O\left(\frac{1}{u^{n}}\right)~~(i\leq m-1)
\end{equation} 
%
%===================================%. 
immediately shows us that Eq.~(\ref{um-eq}) holds. 

From Eq.~(\ref{aa-eq}), $A^{(m)}$ is written by $U^{(1)}, A^{(1)}, h^{(i)}_{ab}(i<m)$ and the 
trace part of $h^{(m)}_{ab}$. 
Since the trace part of $h^{(m)}_{ab}$ is written by $U^{(1)}, A^{(1)}$
and $h^{(i)}_{ab}$ ($i<m$)  
through Eq.~(\ref{rr-eq}), 
$A^{(m)}$ is written only by $U^{(1)}, A^{(1)}, h^{(i)}_{ab}~(i<m)$ in the end. 
Then, by the assumption of the induction, Eq.~(\ref{am-eq}) holds.
Next, the evolution equation for $h^{(m)}_{ab}$ is described by Eq.~(\ref{ab-eq}). 
Expanding Eq.~(\ref{ab-eq}) near the event horizon
and multiplying, $(\mathcal{L}_{n})^{m-1}$, Eq.~(\ref{ab-eq}) becomes the following form on the event horizon
%============<Equation>=============%
%
\begin{equation}
\dot{h}^{(m)}_{ab}+\frac{1}{2}A^{(1)}h^{(m)}_{ab}=F_{ab} ,\label{hm-ev}
\end{equation} 
%
%===================================%
where $F_{ab}$ is a function of $h^{(i)}_{ab}\,(i\leq m)$, 
$\dot{h}^{(i)}_{ab}\,(i<m)$ and so on. 
Acting $\mathcal{L}_{l}$ to Eq.~(\ref{hm-ev}), then, we can see 
that $(1/2)A^{(1)} \dot{h}^{(m)}_{ab} = \dot F_{ab} \sim
\dot{h}^{(j)}_{ab}=O(1/u^{n})$ for $j<m$. 
Thus Eq.~(\ref{hm-eq}) holds. 

Now we can confirm that the induction loop is closed. Then 
we can show that Eq.~(\ref{rigid2}), 
equivalently Eq.~(\ref{rigid}) holds if the spacetime is real analytic. 
Therefore we can show that the null geodesic generator $l$ of the event horizon is the asymptotic Killing vector
at late time in the sense of Eq.~(\ref{rigid}).

\section{Summary and discussion}

We have confirmed that the expansion and shear of the event horizon should decay at late time in the 
vacuum spacetimes. 
Then, assuming the compactness of the cross sections of the event horizon, 
the null geodesic generators on the horizon give us an asymptotic Killing vector $l$ at late time. 
This means that the horizon has late-time symmetry. 
By solving the Einstein equations, then, 
we have found that this late-time symmetry can be extended outside of the black holes. 
Therefore, at late time, there is the asymptotic symmetry outside of black holes. 

If the black hole rotates and there is another asymptotic Killing vector at late time, $k$, which will be 
a stationary Killing vector at a far distance and spacelike near the horizon, $k-l$ is also 
an asymptotic Killing vector expected to correspond to axisymmetry. In this sense, 
we would expect that the rigidity holds in gravitational collapse at late time. 
In these discussions, we assume the compactness of the event horizon. Thus this result 
cannot be applied to other null hypersurfaces which do not have a compact cross section.

There is a remaining issue. In the proof of the rigidity theorem, the exact stationarity does 
show that the null geodesic generators of the horizon are Killing orbits. On the other hand, 
our argument could show us that the null geodesic generators of the horizon is a  Killing orbit 
without assuming the presence of asymptotically stationary Killing vectors. It is likely that 
this difference suggests the existence of important and new points in black hole physics. 

\section*{Acknowledgment}
KT is supported by JSPS Grant-in-Aid for Scientific Research (No.~21-2105).  
TS is supported by
Grant-in-Aid for Scientific Research from the Ministry of Education, Science,
Sports and Culture of Japan (No.~21244033, No.~21111006, No.~20540258, and No.~19GS0219). 
This work is also supported by the Grant-in-Aid for the Global 
COE Program ``The Next Generation of Physics, Spun from Universality 
and Emergence'' from the Ministry of Education, Culture, Sports, Science 
and Technology (MEXT) of Japan.

\appendix

\section{Einstein equations near event horizon}

In Appendix A, using the suitable variables, we will 
show the derivations of the Einstein equations near the event horizon.

\subsection{$(n+1)$ decomposition}

First we describe the formula of the $(n+1)$ decomposition. 
The metric can be written as
%============<Equation>=============%
%
\begin{equation}
g_{AB}\,=\,\epsilon n_{A}n_{B}+\gamma_{AB},
\end{equation} 
%
%===================================%
where $\gamma_{AB}$ is an $n$-dimensional induced metric. 
$n_{A}$ is the unit normal vector with $n_{A}n^{A}=\epsilon=+1$ ($n^A$: spacelike)
or $-1$ ($n^A$: timelike). 

We define the extrinsic curvature as 
%============<Equation>=============%
%
\begin{equation}
K_{AB}\,=\,\frac{1}{2}\mathcal{L}_{n}\gamma_{AB}.
\end{equation} 
%
%===================================%
Now $n_{A}$ can be written as $n_{A}=\epsilon N(d\Omega)_{A}$, where $\Omega$ is 
a function which describes the hypersurface as $\Omega=\text{const.}$ and 
$N$ is the ``lapse'' function. Then the Riemann tensor is decomposed into 
%============<Equation>=============%
%
\begin{equation}
R_{EFGH}\gamma_{A}{}^{E}\gamma_{B}{}^{F}\gamma_{C}{}^{G}\gamma_{D}{}^{H}
   = {}^{(\gamma)}R_{ABCD} - \epsilon K_{AC}K_{BD} + \epsilon K_{AD}K_{BC},
\end{equation} 
%
%===================================%
%============<Equation>=============%
%
\begin{equation}
 R_{EFGD}\gamma_{A}{}^{E}\gamma_{B}{}^{F}\gamma_{C}{}^{G} n^{D}
   = \nabla_{A}K_{BC} - \nabla_{B}K_{AC},
\end{equation} 
%
%===================================%%============<Equation>=============%
%
\begin{equation}
R_{ACBD}n^{C}n^{D} = - \mathcal{L}_{n} K_{AB} + K_{AC}K_{B}{}^{C} 
   - \epsilon \frac{1}{N}\nabla_{A}\nabla_{B} N,
\end{equation} 
%
%===================================%
where $\nabla_{A}$ denotes the covariant derivative with respect to $\gamma_{AB}$.

The Ricci tensor is decomposed into 
%============<Equation>=============%
%
\begin{equation}
  R_{AB}n^{A}n^{B} = - \mathcal{L}_{n}K - K_{AB}K^{AB}-\epsilon \frac{1}{N}\nabla^{2}N,
\end{equation}
%
%===================================%
%============<Equation>=============%
%
\begin{equation}
  R_{AC}n^{A}\gamma_{B}{}^{C} = \nabla^{A}K_{AB} - \nabla_{B}K,
\end{equation}
%
%===================================%
%============<Equation>=============%
%
\begin{equation}
  R_{CD}\gamma_{A}{}^{C}\gamma_{B}{}^{D} = 
   {}^{(\gamma)}R_{AB} - \epsilon\mathcal{L}_{n} K_{AB}
   - \epsilon KK_{AB} + 2\epsilon K_{AC}K_{B}{}^{C}
   - \frac{1}{N}\nabla_{A}\nabla_{B}N.
\end{equation}
%
%===================================%

The Ricci scalar is written as 
%============<Equation>=============%
%
\begin{equation}
  \begin{aligned}
   R =& {}^{(\gamma)}R - 2\epsilon \mathcal{L}_{n}K - \epsilon K^{2}
   - \epsilon K_{AB}K^{AB} - \frac{2}{N}\nabla^{2}N\\
   =& {}^{(\gamma)}R + \epsilon K^{2}
   - \epsilon K_{AB}K^{AB} - \frac{2}{N}\nabla^{2}N - 2\epsilon\nabla_{A}(Kn^{A}).
  \end{aligned}
\end{equation}
%
%===================================%

The components of the Einstein tensor are 
%============<Equation>=============%
%
\begin{equation}
  G_{AB}n^{A}n^{B} = \frac{1}{2}
   (- \epsilon {}^{(\gamma)}R + K^2 - K_{AB}K^{AB}),
 \end{equation}
 \begin{equation}
  G_{AC}n^{A}\gamma_{B}{}^{C} = \nabla^{A} K_{AB} - \nabla_{B} K,
 \end{equation}
 \begin{equation}
  \begin{aligned}
   G_{CD}\gamma_{A}{}^{C}\gamma_{B}{}^{D} =& 
   {}^{(\gamma)}G_{AB}
   - \epsilon KK_{AB} + 2\epsilon K_{AC}K_{B}{}^{C}
   + \frac{\epsilon}{2}\gamma_{AB}(K_{CD}K^{CD} + K^{2})
   \\
   & - \epsilon\mathcal{L}_{n} K_{AB} + \epsilon\gamma_{AB}\mathcal{L}_{n}K
   - \frac{1}{N}\nabla_{A}\nabla_{B}N + \frac{1}{N}\gamma_{AB}\nabla^{2}N. 
  \end{aligned}
 \end{equation}
%
%===================================%

\subsection{Einstein equations}

We apply the $(n+1)$ decomposition presented in the previous section 
to the $r$-constant surface in our current  four dimensional metric form: 
%============<Equation>=============%
%
\begin{equation}
ds^{2}\,=\,-Adu^{2}+2dudr+h_{ab}(dx^{a}+U^{a}du)(dx^{b}+U^{b}du).
\end{equation} 
%
%===================================%
We express the above in the following form
%============<Equation>=============%
%
\begin{equation}
ds^{2}\,=\,N^{2}dr^{2}+q_{\mu\nu}(dx^{\mu}+N^{\mu}dr)(dx^{\nu}+N^{\nu}dr),
\end{equation} 
%
%===================================%
where
%============<Equation>=============%
%
\begin{equation}
N^{2}\,=\,\frac{1}{A},
\end{equation} 
\begin{equation}
N^{u}\,=\,-\frac{1}{A},
\end{equation} 
\begin{equation}
N^{a}\,=\,\frac{1}{A}U^{a}.
\end{equation} 
%
%===================================%
$q_{\mu\nu}$ is the induced metric on the $r$-const. hypersurface as
%============<Equation>=============%
%
\begin{eqnarray}
 q_{\mu\nu} = 
   \begin{pmatrix}
    - A + U^{a}U_{a} & U_{b}\\
    U_{a} & h_{ab}
   \end{pmatrix}.
\end{eqnarray}
%
%===================================%
Note that the Latin indices $a,b,...$ and the Greek 
indices $\mu,\nu,..$ are raised and lowered by $h_{ab}$ and $q_{\mu\nu}$
respectively. The unit normal vector to the $r$-const. hypersurface is 
given by $m_{A}=N(dr)_{A}$ and $m^{A}=N^{-1}(\partial_{r}-N^{\mu}\partial_{\mu})^A$. 

The extrinsic curvature on the $r$-const. hypersurface is defined as 
%============<Equation>=============%
%
\begin{equation}
K_{\mu\nu}\,=\,\frac{1}{2}\mathcal{L}_{m}q_{\mu\nu}\,=\,
\frac{1}{2N}(\partial_{r}q_{\mu\nu}-\mathcal{D}_{\mu}N_{\nu}-\mathcal{D}_{\nu}N_{\mu}),
\end{equation} 
%
%===================================%
where $\mathcal{D}_{\mu}$ is the covariant derivative with respect to $q_{\mu\nu}$. 

We rewrite the induced metric as
%============<Equation>=============%
%
\begin{equation}
q_{\mu\nu}dx^{\mu}dx^\nu=-\alpha^{2}du^{2}+h_{ab}(dx^{a}+\beta^{a}du)(dx^{b}+\beta^{b}du),
\end{equation} 
%
%===================================%
where
%============<Equation>=============%
%
\begin{equation}
\alpha^{2}\,=\,A\,,\,\beta^{a}\,=\,U^{a}.
\end{equation} 
%
%===================================%
The timelike unit vector to the $u$-const. surface is given by $u=-\alpha du$ 
and $u^{\mu}=\alpha^{-1}(\partial_{u}-\beta^{a}
\partial_{a})^{\mu}$.

The extrinsic curvature on the $u$-const. surface is given by 
%============<Equation>=============%
%
\begin{equation}
k_{ab}\,=\,\frac{1}{2}\mathcal{L}_{u}h_{ab}\,=\,\frac{1}{2\alpha}(\partial_{u}h_{ab}-
\bar{D}_{a}\beta_{b}-\bar{D}_{b}\beta_{a}),
\end{equation} 
%
%===================================%
where $\bar{D}_{a}$ is the covariant derivative with respect to $h_{ab}$. 

We introduce the following quantities for later convenience 
%============<Equation>=============%
%
\begin{equation}
\Theta\equiv K_{\,u\nu}u^{\mu}u^{\nu}\,=\,-\frac{1}{N}\partial_{r}(\log{\alpha})-\mathcal{L}_{u}\log{N},
\end{equation} 
\begin{equation}
\rho^{a}\equiv K^{a}_{\mu}u^{\mu}\,=\,\frac{1}{2}\partial_{r}\beta^{a}+\bar{D}^{a}\log{\alpha},
\end{equation} 
\begin{equation}
\kappa_{ab}\equiv K_{cd}h_{a}{}^{c}h_{b}{}^{d}\,=\,\frac{\alpha}{2}\partial_{r}h_{ab}+k_{ab},
\end{equation} 
%
%===================================%
and 
%============<Equation>=============%
%
\begin{equation}
\kappa\,=\, \kappa_{ab}h^{ab}\,=\,\frac{\alpha}{2}(\log{h})^{'}+k,
\end{equation} 
%
%===================================%
where $\rho_{\mu}u^{\mu}=0=\kappa_{\mu\nu}u^{\mu}$, $h=\det{h}_{ab}$ and $k=k_{ab}h^{ab}$.
The prime denotes the $r$ differentiation. Then $K_{\mu\nu}$ can be written as 
%============<Equation>=============%
%
\begin{equation}
K_{\mu\nu}\,=\,\Theta u_{\mu}u_{\nu} -2\rho_{(\mu}u_{\nu )}+\kappa_{\mu\nu}.
\end{equation} 
%
%===================================%

Using these quantities, we can decompose the  four dimensional Ricci tensor $R_{AB}$ into the quantities 
on two dimensional space:
%============<Equation>=============%
%
\begin{eqnarray}
R_{AB}m^{A}m^{B}&=&\frac{1}{N}(\Theta-\kappa)^{'}+\mathcal{L}_{u}(\Theta-\kappa)-\Theta^{2}+2\rho_{a}\rho^{a}
-\kappa_{ab}\kappa^{ab} \notag \\
&&~~~~~~~~~~~~~~~
+\frac{1}{N}(\mathcal{L}_{u}\mathcal{L}_{u}N + k\mathcal{L}_{u}N
 - \bar{D}^{2}N - \bar{D}^{a}N\bar{D}_{a}\log{\alpha}), \label{rr-eq2}
\end{eqnarray}
\begin{equation}
R_{AB}m^{A}q^{B}{}_{C}u^{C}\,=\,-\mathcal{L}_{u}\kappa+\bar{D}^{a}\rho_{a}-k_{ab}\kappa^{ab}+2\rho^{a}\bar{D}_{a}\log{\alpha}
-\Theta k, \label{ru-eq}
\end{equation} 
\begin{eqnarray}
 R_{AB}q^{A}{}_{C}q^{B}{}_{D}u^{C}u^{D}&=&-\frac{1}{N}\Theta^{'}-\mathcal{L}_{u}\Theta +\Theta^{2}-\Theta\kappa
 -2\rho^{a}\rho_{a}-2\rho^{a}\bar{D}_{a}\log{\frac{N}{\alpha}}\notag \\
&&~~~~~~~~~~~~~~~~
+\bar{D}^{a}\log{\alpha}\bar{D}_{a}\log{N}
+\frac{1}{\alpha}\bar{D}^{2}\alpha -\mathcal{L}_{u}k -k_{ab}k^{ab}-\frac{1}{N}\mathcal{L}_{u}\mathcal{L}_{u}N, \label{uu-eq}
\end{eqnarray} 
\begin{equation}
R_{AB}m^{A}h^{Ba}\,=\,\Theta\bar{D}^{a}\log{\alpha}-2\rho_{b}k^{ab}-k\rho^{a}+\bar{D}_{b}\kappa^{ab}-\bar{D}^{a}\kappa
+\bar{D}^{a}\Theta +\kappa^{ab}\bar{D}_{b}\log{\alpha}-\frac{1}{\alpha}(\partial_{u}\rho^{a}-\mathcal{L}_{\beta}\rho^{a}), \label{ra-eq2}
\end{equation} 
\begin{eqnarray}
R_{AB}q^{A}{}_{C}u^{C}h^{Bb}&=&\bar{D}_{a}k^{ab}-\bar{D}^{b}k
 -\rho^{b}\kappa
 - 2\rho_{a}\kappa^{ab}
-\frac{1}{N}\bar{D}^{b}\mathcal{L}_{u}N +k^{ab}\bar{D}_{a}\log{N} -\frac{1}{N}(\rho^{b})^{'}\notag \\
&&~~~~~~~~~~~
 - \Theta\bar{D}^{b}\log{\frac{N}{\alpha}}-\kappa^{ab}\bar{D}_{a}\log{\frac{N}{\alpha}}
-\frac{1}{\alpha}(\partial_{u}\rho^{b}-\mathcal{L}_{\beta}\rho^{b}), \label{ua-eq}
\end{eqnarray}
and
\begin{eqnarray}
R_{AB}h^{A}{}_{a}h^{B}{}_{b}\,=\,{}^{(h)}R_{ab}+\mathcal{L}_{u}k_{ab}+kk_{ab}&-&2k_{ac}k_{b}{}^{c}-\frac{1}{\alpha}\bar{D}_{a}\bar{D}_{b}\alpha
-\frac{1}{N}\kappa_{ab}^{'}-\frac{1}{\alpha}\dot{\kappa}_{ab}+\frac{1}{\alpha}\mathcal{L}_{\beta}\kappa_{ab}
- 2\rho_{(a}\bar{D}_{b)}\log{\frac{N}{\alpha}} \notag \\
 &+&(\Theta-\kappa)\kappa_{ab} -2\rho_{a}\rho_{b} +2\kappa_{ac}\kappa_{b}{}^{c}-\frac{1}{N}\bar{D}_{a}\bar{D}_{b}N+k_{ab}\mathcal{L}_{u}\log{N}, \label{ab-eq2}
\end{eqnarray}
%
%===================================%
where the dot denotes $\partial_{u}$. The vacuum Einstein equation is given by $R_{AB}=0$.

\section{Explicit form of Einstein equations}

In Appendix B, 
we describe the components of the Einstein equations in terms of the metric functions explicitly.

Using 
%============<Equation>=============%
%
\begin{equation}
u=\alpha^{-1}(l-U^{a}\partial_{a})=\alpha^{-1}(\partial_u-U^{a}\partial_{a}),
\end{equation} 
\begin{equation}
\Theta\,=\,-(A^{1/2})^{'} +\frac{1}{2}\mathcal{L}_{u}\log{A},
\end{equation} 
\begin{equation}
\rho^{a}\,=\,\frac{1}{2}U^{a}{}^{'}+\frac{1}{2}\bar{D}^{a}\log{A},
\end{equation} 
and
\begin{equation}
\kappa_{ab}\,=\,\frac{A^{1/2}}{2}h^{'}_{ab}+\frac{1}{2A^{1/2}}(\dot{h}_{ab}-\bar{D}_{a}U_{b}-\bar{D}_{b}U_{a}),
\end{equation}
%
%===================================%
we can rewrite the Einstein equation $R_{AB}=0$ in terms of our metric form. 
We will not provide all components of the Einstein equation explicitly. 

For example, $R_{AB}h^A{}_a h^B{}_b=0$ becomes
%============<Equation>=============%
%
\begin{eqnarray}
\mathcal{L}_{l}h^{'}_{ab}+\frac{1}{2}A^{'}h^{'}_{ab}&=&{}^{(h)}R_{ab}+\frac{A}{2}h^{cd}h^{'}_{ac}h^{'}_{bd}+\frac{1}{2}h^{cd}(h^{'}_{ac}\dot{h}_{bd}
+h^{'}_{bd}\dot{h}_{ac})-\frac{1}{2}h^{'}_{ac}(\bar{D}_{b}U^{c}+\bar{D}^{c}U_{b}) 
+ \mathcal L_U h'_{ab}
\notag \\
&&~~~~~
-\frac{1}{2}h^{'}_{bc}(\bar{D}_{a}U^{c}+\bar{D}^{c}U_{a})
-\frac{1}{2}h_{ac}h_{bd}U^{c}{}^{'}U^{d}{}^{'}
 -\frac{1}{4}[\dot{(\log{h})}-2\bar{D}_{a}U^{a}]h^{'}_{ab} \notag \\
&&~~~~~
-\frac{1}{4}(\log{h})^{'}
(A h^{'}_{ab} +\dot{h}_{ab} -\bar{D}_{a}U_{b}-\bar{D}_{b}U_{a})
-\frac{1}{2}Ah^{''}_{ab}+\frac{1}{2}(h_{bc}\bar{D}_{a}U^{c}{}^{'}+h_{ac}\bar{D}_{b}U^{c}{}^{'}).
\end{eqnarray} 
%
%===================================%
This determines the evolutions of $h_{ab}$. For $R_{AB}m^{A}h^{B a}=0$, we have 
%============<Equation>=============%
%
\begin{eqnarray}
\mathcal{L}_{l}U^{a}{}^{'}&=&-\bar{D}^{a}A^{'}-h^{ac} U^{b}{}^{'} (\dot{h}_{bc}-\bar{D}_{a}U_{b}-\bar{D}_{b}U_{a})
+h^{ab}\bar{D}^{c}(Ah^{'}_{bc}+\dot{h}_{bc}-\bar{D}_{b}U_{c}-\bar{D}_{c}U_{b})-A\bar{D}^{a}(\log{h})^{'} \notag \\
&&~~~~-\frac{1}{2}U^{a}{}^{'}[\dot{(\log{h})}-2\bar{D_{a}}U^{a}]
-\frac{1}{2}(\log{h})^{'}\bar{D}^{a}A-\bar{D}^{a}[\dot{(\log{h})}-2\bar{D}_{b}U^{b}]
+ \mathcal{L}_{U} U^{a}{}^{'} \label{u-eq}.
\end{eqnarray}
%
%===================================%

\section{The gauge issue for $A^{(1)}$ [Eq.~(\ref{a1-eq3})]}

In Appendix C we will show the presence of the gauge where Eq.~(\ref{a1-eq3}) is satisfied. 
In our coordinate $x^{A}=(u,r,x^{a})$, the metric can be written as
%============<Equation>=============%
%
\begin{equation}
ds^{2}\,=\,-Adu^{2}+2dudr +h_{ab}(dx^{a}+U^{a}du)(dx^{b}+U^{b}du). 
\end{equation} 
%
%===================================%
The components of the metric are expanded near the event horizon $(r=0)$ as
%============<Equation>=============%
%
\begin{equation}
A\,=\,rA^{(1)}+O(r^{2}), \label{a1-exp}
\end{equation} 
\begin{equation}
U^{a}\,=\,rU^{(1)a}+O(r^{2}) \label{u1-exp}
\end{equation}
and
\begin{equation}
h_{ab}\,=\,h^{(0)}_{ab}+O(r).
\end{equation}
%
%===================================%
Here $A^{(1)}$ can be decomposed as
%============<Equation>=============%
%
\begin{equation}
A^{(1)}\,=\,A_{0}^{(1)}+\tilde{A}^{(1)}(u,x^{a}), \label{a1-exp2}
\end{equation} 
%
%===================================%
where $A_{0}^{(1)}$ is set to be a constant as shown in Ref. \cite{Hollands:2006rj}. 

When we consider the gauge transformation $x^{A}\rightarrow x^{A}+\xi^{A}$,
the metric is transformed as
%============<Equation>=============%
%
\begin{equation}
g_{AB}\rightarrow g_{AB}+
\mathcal L_\xi g_{AB}
\equiv g_{AB}+\delta g_{AB}.
\end{equation} 
%
%===================================%
To keep our gauge, the following conditions will be imposed:
%============<Equation>=============%
%
\begin{gather}
\delta g_{ur}\,=\,0\,,\,\delta g_{rr}\,=\,0\,,\,\delta g_{ra}\,=\,0\,, \notag \\
\delta g_{uu}\,=\,O(r)\,,\,\delta g_{ua}\,=\,O(r)\,,\,\delta g_{ab}\,=\,O(1). \label{gcond}
\end{gather} 
From $\delta g_{rr} = 2\partial_r \xi^u = 0$, we have 
$\xi^u = f(u,x^a)$.
%
%===================================%
Since $\delta g_{ra}$ and $\delta g_{ur}$ are given by  
%============<Equation>=============%
%
\begin{equation}
\delta g_{ra}\,=\,\partial_a \xi^u + U_a\partial_r \xi^u +
 h_{ab}\partial_r \xi^b,
\end{equation}
\begin{equation}
\delta g_{ur}\,=\,(-A+U^{a}U_{a})\partial_{r}\xi^u + \partial_{r}\xi^r +
 \partial_{u}\xi^u + U_a \partial_r \xi^a,
\end{equation}
%
%===================================%
$\delta g_{ra}=0$ and $\delta g_{ur}=0$ lead to  
%============<Equation>=============%
%
\begin{equation}
 \xi^r = - r\partial_u f + \partial_a f \int^r U^a dr', \quad
  \xi^a = - \partial_b f \int^r h^{ab} dr'.
\end{equation}
%
%===================================%
Then $\delta g_{uu}$ becomes 
%============<Equation>=============%
%
\begin{equation}
\delta g_{uu}\,=\,r[ -\partial_{u}(fA^{(1)})-\partial^{2}_{u}f ] +O(r^{2}),
\end{equation} 
%
%===================================%
where we used Eqs.~(\ref{a1-exp}) and (\ref{u1-exp}). 
This means that $A^{(1)}$ is transformed under the gauge transformation as
%============<Equation>=============%
%
\begin{equation}
A^{(1)}\rightarrow A^{(1)}+\partial_{u}(fA^{(1)})+\partial_{u}^{2}f.
\end{equation} 
%
%===================================%
Thus if we choose $f$ as 
%============<Equation>=============%
%
\begin{equation}
f\,=\,-\frac{1}{A^{(1)}_{0}}\int^{u}du^{'}\tilde{A}^{(1)},
\end{equation} 
%
%===================================% 
$A^{(1)}$ is transformed as 
%============<Equation>=============%
%
\begin{equation}
A^{(1)}\rightarrow \bar A^{(1)}\,=\,A^{(1)}_{0}-\frac{1}{A^{(1)}_{0}}[\partial_{u}\tilde{A}^{(1)}
+\partial_{u}(f\tilde{A}^{(1)})].
\end{equation} 
%
%===================================%

Let assume that $\tilde A^{(1)}$ decays as $u \to \infty$. For the moment, 
we write $\tilde{A}^{(1)}=O(1/u^{m})$, where 
$m$ is an integer. If $m \geq n$, Eq.~(\ref{a1-eq3}) is already satisfied. Therefore, we 
suppose that $m$ is smaller than $n$. 

In the current gauge transformation, the transformed $\bar A^{(1)}$ satisfies
%============<Equation>=============%
%
\begin{equation}
\bar A^{(1)}\,=\,A^{(1)}_{0}+O(1/u^{m+1}). 
\end{equation} 
%
%===================================%
Repeating this procedure, we can always choose the gauge satisfying 
%============<Equation>=============%
%
\begin{equation}
\mathcal{L}_{l}A^{(1)}\,=\,O\left(\frac{1}{u^{n+1}}\right). 
\end{equation} 
%
%===================================%
If one wishes, one can continue the same procedure and then achieve an arbitrary 
faster decaying rate. But, the above is enough for our current purpose.


\begin{thebibliography}{99}

\bibitem{Israel}
W.~Israel, Phys.\ Rev.\ {\bf 164}, 1776 (1967);
B.~Carter, Phys.\ Rev.\ Lett.\ {\bf 26}, 331 (1971);
S.~W.~Hawking, Commun.\ Math.\ Phys.\ {\bf 25}, 152 (1972);
D.~C.~Robinson, Phys.\ Rev.\ Lett.\ {\bf 34}, 905 (1975);
P.~O.~Mazur, J.\ Phys.\ {\bf A15}, 3173 (1982); 
For a review, see M. Heusler, {\it Black Hole Uniqueness Theorems}, (Cambridge University Press, London, 1996);
P.~O.~Mazur, hep-th/0101012; 
G.~L.~Bunting, PhD thesis, Univ.\ of New England, Armidale (1983). 

%\cite{Hawking:1973uf}
\bibitem{Hawking:1973uf}
  S.~W.~Hawking and G.~F.~R.~Ellis, {\it The Large scale structure of space-time},
(Cambridge Univ. Press, Cambridge, 1973). 
  
%\cite{Hawking:1971vc}
\bibitem{Hawking:1971vc}
  S.~W.~Hawking,
  %``Black holes in general relativity,''
  Commun.\ Math.\ Phys.\  {\bf 25}, 152-166 (1972).

%\cite{Friedrich:1998wq}
\bibitem{Friedrich:1998wq}
  H.~Friedrich, I.~Racz, R.~M.~Wald,
  %``On the rigidity theorem for space-times with a stationary event horizon or a compact Cauchy horizon,''
  Commun.\ Math.\ Phys.\  {\bf 204}, 691-707 (1999).
%  [gr-qc/9811021].
  
%\cite{Racz:1999ne}
\bibitem{Racz:1999ne}
  I.~Racz,
  %``On further generalization of the rigidity theorem for space-times with a stationary event horizon or a compact Cauchy horizon,''
  Class.\ Quant.\ Grav.\  {\bf 17}, 153-178 (2000).
%  [gr-qc/9901029].

%\cite{Gundlach:1993tp}
\bibitem{Gundlach:1993tp}
  C.~Gundlach, R.~H.~Price, J.~Pullin,
  %``Late time behavior of stellar collapse and explosions: 1. Linearized perturbations,''
  Phys.\ Rev.\  {\bf D49}, 883-889 (1994).
%  [gr-qc/9307009].

%\cite{Barack:1999ma}
\bibitem{Barack:1999ma}
  L.~Barack, A.~Ori,
  %``Late time decay of scalar perturbations outside rotating black holes,''
  Phys.\ Rev.\ Lett.\  {\bf 82}, 4388 (1999).
%  [gr-qc/9902082].
  
%\cite{Barack:1999ya}
\bibitem{Barack:1999ya}
  L.~Barack, A.~Ori,
  %``Late time decay of gravitational and electromagnetic perturbations along the event horizon,''
  Phys.\ Rev.\  {\bf D60}, 124005 (1999).
%  [gr-qc/9907085].

%\cite{Barack:1998bw}
%\bibitem{Barack:1998bw}
%  L.~Barack,
%  %``Late time dynamics of scalar perturbations outside black holes. 2. Schwarzschild geometry,''
%  Phys.\ Rev.\  {\bf D59}, 044017 (1999).
%  [gr-qc/9811028].

%\cite{Tanabe:2011es}
\bibitem{Tanabe2011}
  K.~Tanabe, S.~Kinoshita, T.~Shiromizu,
  %``Asymptotic flatness at null infinity in arbitrary dimensions,''
  Phys.\ Rev.\  {\bf D84}, 044055 (2011).
%  [arXiv:1104.0303 [gr-qc]].



%\cite{Hayward:1993tt}
\bibitem{HSN}
  S.~A.~Hayward, T.~Shiromizu, K.~-i.~Nakao,
  %``A Cosmological constant limits the size of black holes,''
  Phys.\ Rev.\  {\bf D49}, 5080-5085 (1994).
%  [gr-qc/9309004].

%\cite{Booth:2003ji}
\bibitem{Booth:2003ji}
  I.~Booth, S.~Fairhurst,
  %``The First law for slowly evolving horizons,''
  Phys.\ Rev.\ Lett.\  {\bf 92}, 011102 (2004).
%  [gr-qc/0307087].

%\cite{Booth:2006bn}
\bibitem{Booth:2006bn}
  I.~Booth, S.~Fairhurst,
  %``Isolated, slowly evolving, and dynamical trapping horizons: Geometry and mechanics from surface deformations,''
  Phys.\ Rev.\  {\bf D75}, 084019 (2007).
%  [gr-qc/0610032].

%\cite{Ashtekar:2004cn}
\bibitem{Ashtekar:2004cn}
  A.~Ashtekar, B.~Krishnan,
  %``Isolated and dynamical horizons and their applications,''
  Living Rev.\ Rel.\  {\bf 7}, 10 (2004).
%  [gr-qc/0407042].

%\cite{Hollands:2006rj}
\bibitem{Hollands:2006rj}
  S.~Hollands, A.~Ishibashi, R.~M.~Wald,
  %``A Higher dimensional stationary rotating black hole must be axisymmetric,''
  Commun.\ Math.\ Phys.\  {\bf 271}, 699-722 (2007).
%  [gr-qc/0605106].

%\cite{Gen:1998ay}
%\bibitem{Gen:1998ay}
%  U.~Gen, T.~Shiromizu,
%  %``Asymptotically Schwarzschild space-times,''
%  J.\ Math.\ Phys.\  {\bf 40}, 2021-2031 (1999).
%  [gr-qc/9810040].

%\cite{Tanabe:2011fd}
%\bibitem{Tanabe:2011fd}
%  K.~Tanabe, T.~Shiromizu,
%  %``Asymptotic structure at timelike infinity: higher orders,''
%  [arXiv:1103.5183 [gr-qc]].



\end{thebibliography}
\end{document}